\documentclass{moriond}
\usepackage{xspace}

\def\be{\begin{equation}}
\def\ee{\end{equation}}
\def\bea{\begin{eqnarray}}
\def\eea{\end{eqnarray}}

\def\lbpkmumu{\ensuremath{\Lambda_b^0 \to p K^- \mu^+\mu^-}\xspace}
\def\lbbarpkmumu{\ensuremath{\bar \Lambda_b^0 \to \bar p K^+ \mu^+\mu^-}\xspace}
\def\lbpkjpsi{\ensuremath{\Lambda_b^0 \to p K^- J/\psi}\xspace}

\def\lbppimumu{\ensuremath{\Lambda_b^0 \to p \pi^- \mu^+\mu^-}\xspace}

\def\ks{\ifmmode \mr{K^0_S} \else $\mr{K^0_S}$\xspace\fi}
\def\xf{$x_\mr{F}$\xspace}
\def\pt{$p_{\mr{T}}$\xspace}

\def\bsmumu{\ensuremath{B^0_s \to \mu^{+} \mu^{-}}\xspace}
\def\bee{\ensuremath{B^0_{(s)} \to e^{+} e^{-}}\xspace}

\def\bellell{\ensuremath{B^0_{(s)} \to \ell^{+} \ell^{-}}\xspace}

\def\bdsmumu{\ensuremath{B^0_{d,s} \to \mu^{+} \mu^{-}}\xspace}

\def\btosellell{\ensuremath{b \to s \ell^{+} \ell^{-}}\xspace}
\def\btoqellell{\ensuremath{b \to q \ell^{+} \ell^{-}}\xspace}
\def\btosmumu{\ensuremath{b \to s \mu^{+} \mu^{-}}\xspace}
\def\btodellell{\ensuremath{b \to d \ell^{+} \ell^{-}}\xspace}
\def\bdkstarmumu{\ensuremath{B^0_d \to \bar K^{\ast 0} \mu^{+} \mu^{-}}\xspace}
\def\bskstarmumu{\ensuremath{B^0_s \to \bar K^{\ast 0} \mu^{+} \mu^{-}}\xspace}

\def\invfb{\ensuremath{fb^{-1}}\xspace}
\def\bdkstarjpsi{\ensuremath{B^0_d \to \bar K^{\ast 0} J/\psi}\xspace}
\def\bskstarjpsi{\ensuremath{B^0_s \to \bar K^{\ast 0} J/\psi}\xspace}

\def\bdkstarellell{\ensuremath{B^0_d \to \bar K^{\ast 0} \ell^{+} \ell^{-}}\xspace}

\def\bukmumu{\ensuremath{B^+ \to K^+ \mu^{+} \mu^{-}}\xspace}
\def\bdkmumu{\ensuremath{B^0 \to K^0 \mu^{+} \mu^{-}}\xspace}
\def\bukstarmumu{\ensuremath{B^+ \to K^{\ast +} \mu^{+} \mu^{-}}\xspace}

\def\bukellell{\ensuremath{B^+ \to K^+ \ell^{+} \ell^{-}}\xspace}

\def\bsphimumu{\ensuremath{B^0_s \to \phi \mu^- \mu^+}\xspace}

\def\bdstautau{\ensuremath{B^0_{(s)} \to \tau^{+} \tau^{-}}\xspace}

\def\ks{\ensuremath{K^0_{\rm{S}}}\xspace}

\def\xf{\ifmmode x_{\mr{F}} \else $x_{\mr{F}}$\xspace \fi}
\def\pt{\ifmmode p_{\mr{T}} \else $p_{\mr{T}}$\xspace \fi}
\def\dll{\ifmmode \Delta \log \mathcal{L} \else $\Delta \log \mathcal{L}$ \fi}

\def\gev{\ensuremath{\rm{GeV}}\xspace}

\def\ks{\ensuremath{K^0_S}\xspace}

\newcommand{\Photo}{}

\begin{document}
\vspace*{4cm}

\title{Search for new physics in $b \to q \ell \ell$ decays } 

\author{ F. Dettori~\footnote{On behalf of the LHCb collaboration} }

\address{Oliver Lodge Laboratory, University of Liverpool, Liverpool, UK}

\maketitle\abstracts{
Recent results obtained in experiments at the LHC 
in the field of rare $b$-hadron decays are reviewed in this contribution, 
with a focus on $b\to q\ell\ell$ processes. 
A general status is presented as well as recently completed measurements. 
}

\section{Introduction}

The study of flavour-changing neutral-current processes (FCNC) is a sensitive tool to explore 
possible interactions beyond those predicted by the Standard Model~(SM). 
In particular, rare hadron decays having tiny probabilities in the SM, can receive possible New Physics (NP) contributions
at the same level or even larger than the SM.   Probing NP using these techniques is
especially sensitive when the SM predictions are very precise; in these cases, 
we can often probe energy scales orders of magnitude higher than those available 
for the direct production of new states at colliders. 
In addition, rare decays offer the possibility of model-independent tests of physics beyond the SM. 
Finally, historically rare decays have been the laboratory of many particle physics discoveries. 

The description of heavy-quark hadrons rare decays is usually made in terms of effective field theories
where SM and NP contributions are computed in terms of local operators and their Wilson coefficients. 
This allows to put model-independent constraints on possible NP couplings. 
As an example possible new (pseudo)-scalar contributions in \btoqellell processes can be defined as a term in the effective hamiltonian 
proportional to $C_{S (P)} \bar q_L b_L \bar \ell (\gamma_5) \ell$
where $C_{S (P)}$ are the Wilson coefficients, and the overall normalisation is not reported. 
Similarly in what follows it is useful to define \mbox{(axial-)vector} contributions as
$C_{9 (10)} \bar q_L \gamma^\mu b_L \bar \ell \gamma_\mu (\gamma_5) \ell$. 
We do not list them, but additional operators also possibly contribute to \btoqellell transitions
as well as the chirally-flipped versions of the ones defined.

In this contribution we review recent results in rare $b$-hadron decays based on 
quark-level \btosellell and \btodellell transitions with a focus on measurements from 
experiments at the Large Hadron Collider (LHC).

\section{Scalar, pseudoscalar and axial-vector couplings}

NP contributions in the form of scalar, pseudoscalar and axial-vector couplings are best tested 
with \bellell decays. The branching fractions of these decays 
are very well predicted in the SM~\cite{Beneke:2017vpq,PhysRevLett.112.101801}
and thus allow for extremely sensitive tests of NP. 
Due to helicity suppression the \bee decays are too rare to be 
currently probed at branching fractions close to the SM contribution. 
The \bdstautau decays are the less rare but, due to the difficulty of reconstructing 
$\tau$ leptons, especially at hadron colliders, the SM branching fraction is still out of reach. 
The current world best limits on these decays have been recently set by LHCb~\cite{Aaij:2017xqt}. 
Nevertheless it is important to probe these decays, especially in light 
of possible hints of lepton non universality (see C. Langenbruch contribution at this same conference~\cite{langenbruch}). 
The \bdsmumu  decays instead, being cleaner from an experimental point of view, 
are being probed by several experiments. 
The first evidence for the \bsmumu decay, 
came from the LHCb experiment~\cite{Aaij:2012nna}, while the definitive observation 
resulted from the combined analysis of the LHCb and CMS Run 1 measurements~\cite{CMS:2014xfa}.
The current world best measurement comes from the last LHCb measurement~\cite{Aaij:2017vad}, 
where the full Run 1 as well as 1.4~\invfb of integrated luminosity collected in Run 2 was exploited leading 
to the first observation of the \bsmumu decay from a single experiment.
The branching fraction measurement yields 
${\cal B}(B^0_s\to\mu^+\mu^-)=\left(3.0\pm 0.6^{+0.3}_{-0.2}\right)\times 10^{-9}$,
in good agreement with recent SM predictions~\cite{Beneke:2017vpq,PhysRevLett.112.101801} 
as well as with the previous measurements. 
The Run 1 ATLAS search for the same decay is instead in slight tension with these results~\cite{Aaboud:2016ire}, 
reporting only a 1.4$\sigma$ excess over background expectations 
and thus putting an upper limit on the branching fraction at 
${\cal B}(B^0_s\to\mu^+\mu^-) < 3.0 \times 10^{-9}$ at 90\% CL, 
lower than the SM prediction. 
For more details on these and other very rare decays measurements see M. van Veghel 
contribution at this same conference~\cite{vanveghel}.

\section{Vector couplings and a possible anomaly}

While scalar and pseudo-scalar couplings seem to be in good agreement with the SM, 
the same is not true for other couplings, especially vector ones ($C_9$). 
Several small discrepancies, which we review briefly, are accumulating giving an overall tension with the SM. 

\begin{figure}
\includegraphics[width = 0.47\textwidth]{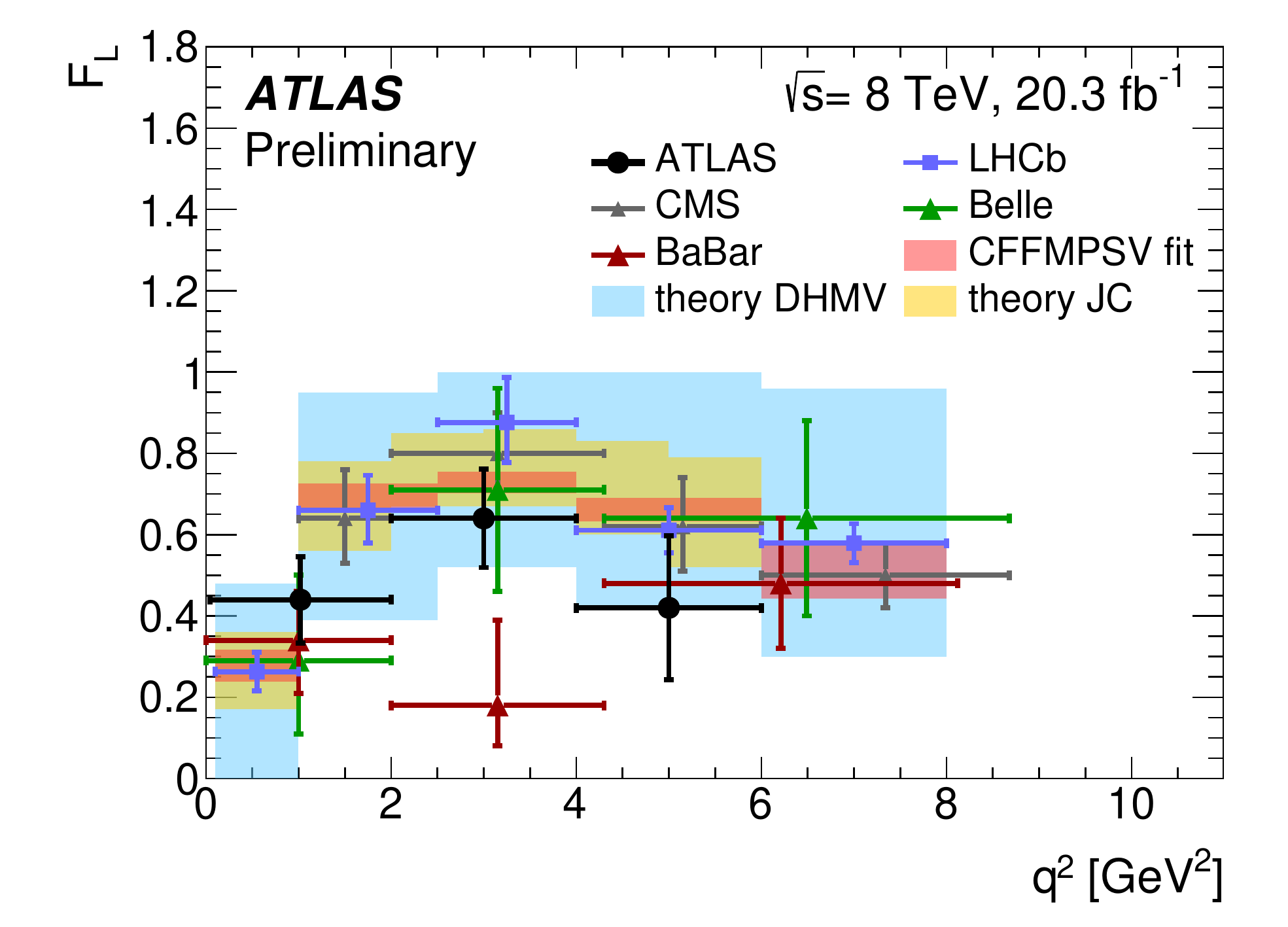}
\includegraphics[width =  0.49\textwidth]{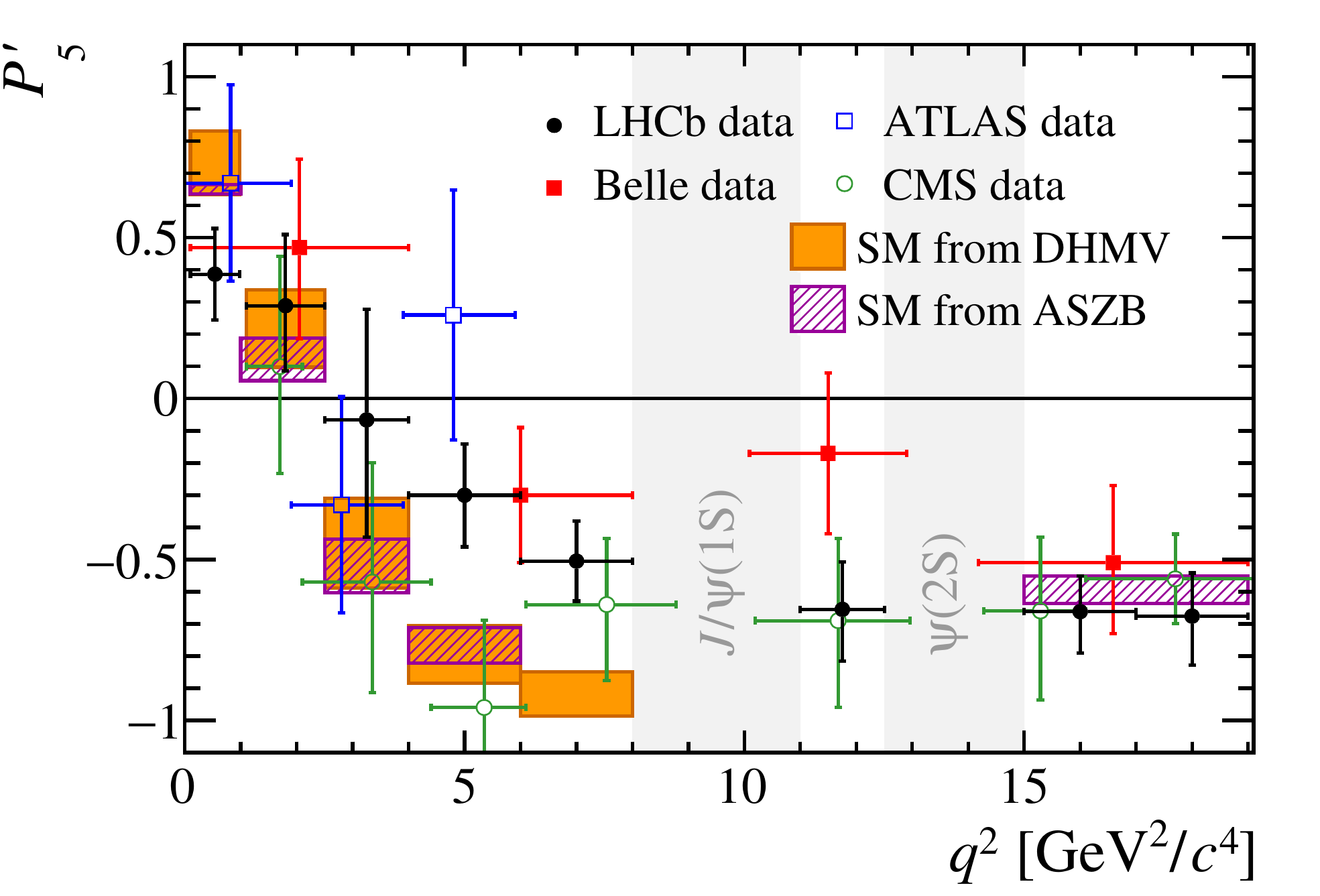}
\caption{Measurements of the $F_L$ and $P^\prime_5$ observables in \bdkstarmumu decays, 
in bins of $q^2$, for different experiments and compared with SM predictions.  }\label{fig:bdkstarmumu}
\end{figure}

We start from the \bdkstarmumu decay; its rich phenomenology allows to test several 
couplings simultaneously through the study of its angular distributions. 
Recent measurements have been done by four experiments, 
often exploiting the so-called optimised observables~\cite{DescotesGenon:2012zf}, 
which have the advantage of reduced theoretical uncertainty. 
Within the Run 1 LHCb measurement~\cite{Aaij:2015oid} 
the full fit to CP-averaged observables shows a tension with the SM with a significance 
of 3.4 standard deviations. 
Considering single variables, the most striking feature is a discrepancy in the $P^\prime_5$ 
parameter. This behaviour seems to be confirmed by ATLAS~\cite{ATLAS:2017dlm}
and Belle~\cite{Wehle:2016yoi} measurements; while the CMS~\cite{CMS:2017ivg} measurement is more in agreement with SM. 
This is shown in Figure~\ref{fig:bdkstarmumu}, where the different measurements are presented
as a function of $q^2$, the dimuon invariant mass squared, and compared with 
SM predictions~\cite{Ciuchini:2015qxb,Descotes-Genon:2014uoa,Jager:2014rwa}.

Similarly, discrepancies are present in the branching fractions of several \btosmumu processes, 
recently measured by LHCb, which tend to show a mild tension with the SM.
In particular the differential and integrated branching fractions of 
\bukmumu, \bukstarmumu, \bdkmumu ~\cite{Aaij:2014pli}  as well as \bdkstarmumu~\cite{Aaij:2016flj} 
and \bsphimumu~\cite{Aaij:2015esa} lie all below the SM prediction, albeit not with great significance. 

These discrepancies summed with the hints of lepton flavour universality
violation in \bukellell, \bdkstarellell and $B\to D^{(\ast)}\ell\nu$ 
decays~\cite{langenbruch} lead to possible global discrepancies with respect to
the SM even above the 5$\sigma$ level. 
In particular, global fits~\cite{nardecchia,Capdevila:2017bsm,Altmannshofer:2017yso,Alok:2017sui}
could indicate possible NP contributions on the $C_9$ Wilson coefficient, 
or to the $C_9$ and $C_{10}$ simultaneously. 
The significance of these discrepancies is almost completely dominated by the
statistical uncertainties, hence they will need to be confirmed
with additional statistics by the same experiments, and cross-checked by other
experiments (such as Belle-II), but give encouraging hints on the direction to follow.

\subsection{Angular analysis of \bukmumu at CMS}

\begin{figure}
 \includegraphics[width = \textwidth]{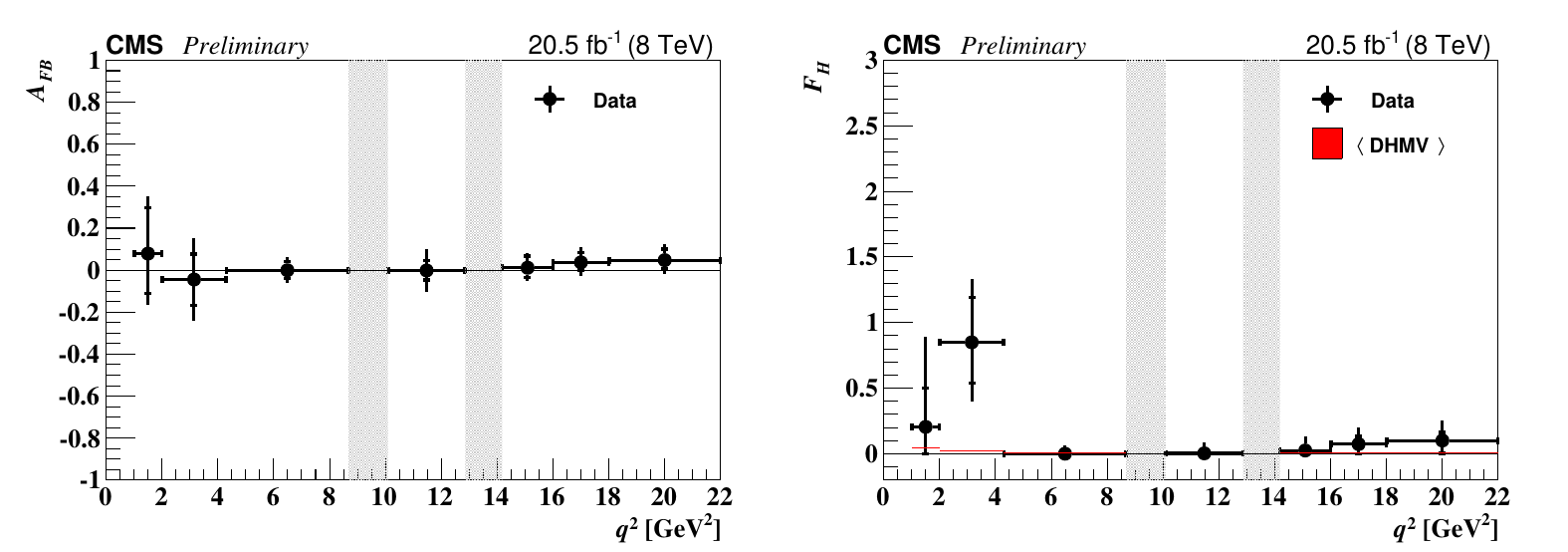}
\caption{Measurement of (left) the $A_{FB}$ parameter and (right) the $F_H$ parameter as a function 
of $q^2$ in the angular analysis of \bukmumu decays in CMS.  }\label{fig:bukmumu}
\end{figure}

Presented for the first time at this conference, the CMS collaboration has performed 
an angular analysis of the \bukmumu decays, exploiting 20.5\invfb of integrated luminosity collected in Run 1
at $\sqrt{s} = 8$ TeV~\cite{CMS-PAS-BPH-15-001}. 
The differential decay rate of this decay can be written as: 
\begin{equation}
\frac{1}{\Gamma_\ell} \frac{d \Gamma_\ell}{d \cos \theta_\ell} = \frac{3}{4}(1 - F_H)(1 - \cos^2 \theta_\ell) + \frac{1}{2}F_H  + A_{FB} \cos \theta_\ell \quad,
\end{equation}
where $\theta_\ell$ is the angle between the direction of the $\mu^-$ and the one of the $K^+$
meson for the $B^+$ decay. The two parameters on which the differential decay rate depends  ($A_{FB}$ and $F_H$)
are a function of $q^2$ and sensitive probes to scalar and tensor new physics contributions. 
The \bukmumu angular analysis was previously performed also by the LHCb~\cite{Aaij:2014tfa}, BaBar~\cite{Aubert:2006vb}, 
 Belle~\cite{Wei:2009zv} and CDF~\cite{Aaltonen:2011ja} experiments. 
The CMS results as a function of $q^2$ are shown in Figure~\ref{fig:bukmumu}, 
as extracted from about 2300 signal candidates. 
The largest systematic uncertainties are due to modelling of the background
components and dominate the total uncertainty in some bins. 
The results are compatible with previous measurements and with SM predictions. 
While this measurement is not yet the world most precise, it is important to
have measurements from different experiments, especially when anomalies such as the ones described
are surfacing. A measurement of the \bukmumu differential branching fraction by CMS is
therefore important and awaited.

\subsection{$b\to d\ell\ell$}

\begin{figure}
\centering
\includegraphics[width = 0.49\textwidth]{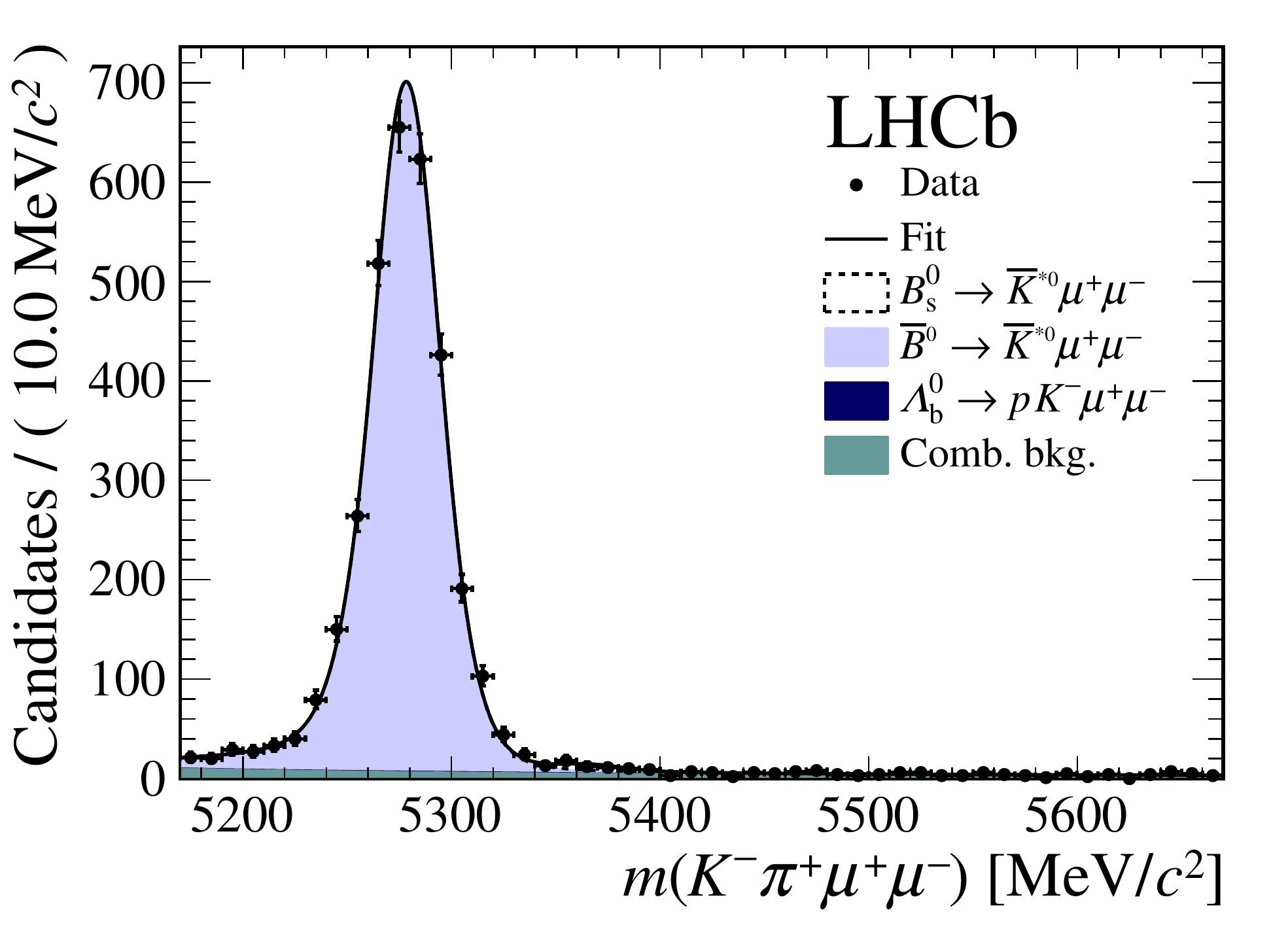}
\includegraphics[width = 0.49\textwidth]{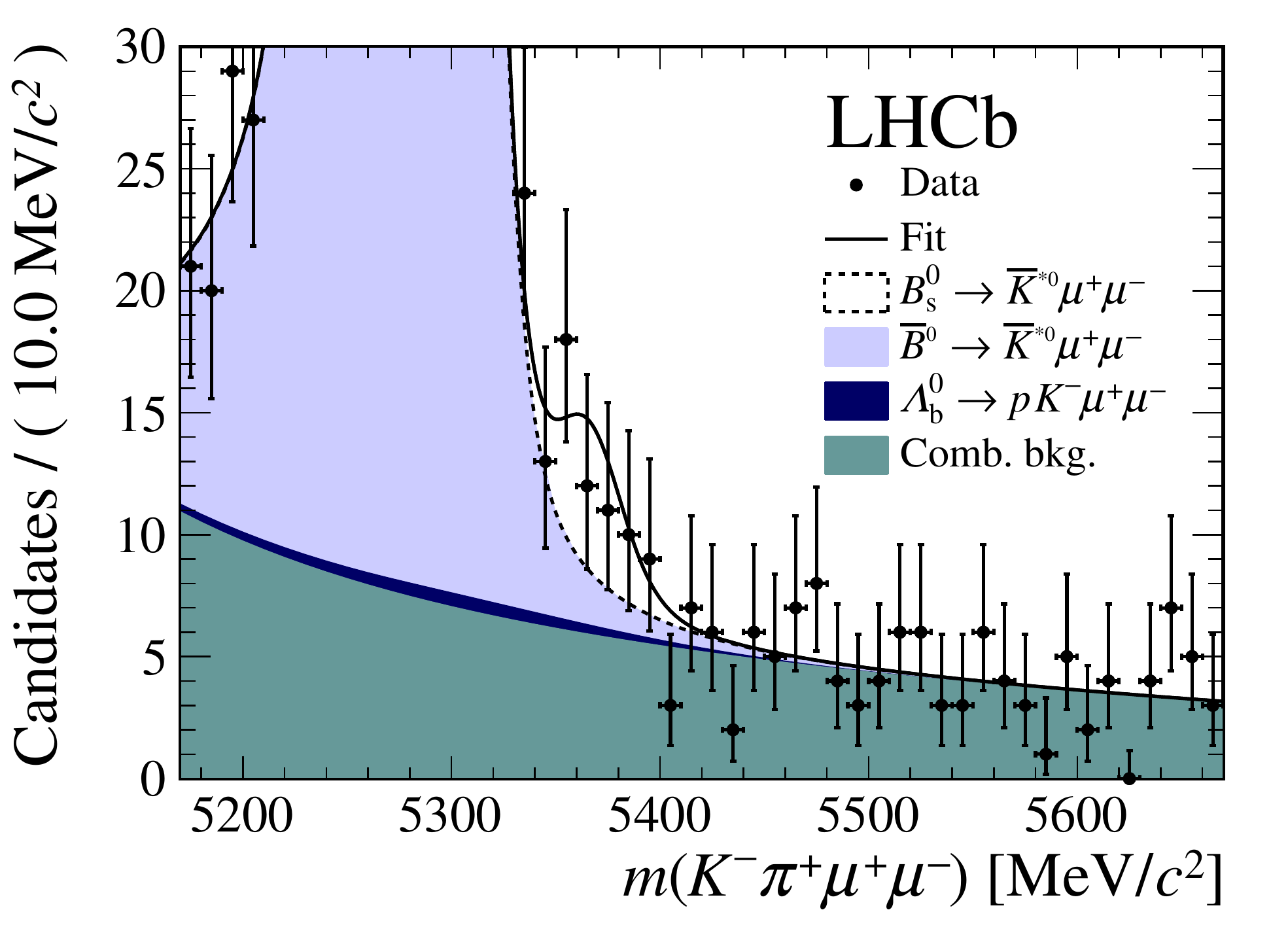}
\caption{Invariant mass distribution for (left) $K^- \pi^+ \mu^+ \mu^-$ 
candidates for the three most sensitive neural network bins of the LHCb analysis 
 and (right) a zoom on the same plot to highlight the \bskstarmumu signal. }\label{fig:bskstarmumu}
\end{figure}

Processes involving \btodellell transitions are even more rare than those with
an $s$ quark due to the additional CKM suppression. 
The observation of these processes was established by LHCb in $B^+\to \pi^+\mu^+
\mu^-$ and $\Lambda_b^0 \to p \pi^- \mu^+\mu^-$ decays~\cite{Aaij:2015nea,Aaij:2017ewm}. 
The \bskstarmumu decay is instead yet to be observed. 
This decay can be sensitive to NP contributions, but is also an important probe
of the $|V_{td}|/|V_{ts}|$ ratio. 
A search for this decay has been performed by LHCb exploiting 4.6\invfb
collected at different energies 
in Run 1 and 2~\cite{Aaij:2018jhg}. 
A common selection for the \bskstarmumu and the control and normalisation
channels \bdkstarmumu, \bdkstarjpsi and \bskstarjpsi is devised. 
The search for the rare non-resonant channel is restricted to 
dimuon invariant masses squared of $0.1 < q^2 < 19.0 ~\gev^2/c^4$
excluding the region $12.5 < q^2 < 15.0 ~\gev^2/c^4$, which is dominated by
charmonium resonances. 
After a loose pre-selection, candidates are classified in bins of a neural
network based 
on geometric and kinematic variables. The background is predominantly composed
of combinatorial background and of the upper tail of the \bdkstarmumu
distribution. 
The distribution of the $K^- \pi^+ \mu^+ \mu^-$ candidates summed over the three
most sensitive bins of neural network
is shown in Figure~\ref{fig:bskstarmumu}. The presence of an excess on top of
the background at the $B^0_s$ mass can be seen. 
From a simultaneous fit a combined significance of 3.4 standard deviations is
derived, including systematic uncertainties, 
which represents the first evidence of the \bskstarmumu decay. 
Normalising the signal to the \bdkstarmumu decay a branching fraction of 
$\mathcal{B}(\bskstarmumu) = (2.9 \pm 1.0 \pm 0.2 \pm 0.3) \times 10^{-8} $ is
obtained, 
where the first uncertainty is statistical, the second systematic and the third
is due to limited knowledge of
parameters used in the normalisation. This result is in agreement with SM
expectations.

\subsection{$b$-baryons}

A different look to \btosellell transitions can be taken by studying $b$-baryon decays, 
the rare decays of which start to be probed only recently. 
In addition to the already mentioned \lbppimumu, 
the LHCb collaboration has studied the \lbpkmumu decay exploiting 
3\invfb of Run 1 data~\cite{Aaij:2017mib}. 
The \lbpkmumu decay is observed for the first time, with large significance. 
Exploiting a signal yield of $600 \pm 44$ candidates, a search for CP violation is performed in this channel
which can be a sensitive probe of NP~\cite{Durieux:2015zwa}.
A measurement of CP violation is built from raw yields as: 
\begin{equation}
\mathcal{A}_{\rm{CP}}\propto \mathcal{A}_{\rm{raw}} = \frac{N(\lbpkmumu) -N(\lbbarpkmumu)}{N(\lbpkmumu) + N(\lbbarpkmumu)}\quad , 
\end{equation}
where the proper conversion to $\mathcal{A}_{\rm{CP}}$ is obtained correcting for production and reconstruction asymmetries. 
For additional robustness, CP violation is searched in the difference of this observable between the \lbpkmumu
and the control channel \lbpkjpsi, on which no significant CP violation is expected. 
This measurement results in: $\Delta\mathcal{A}_{C\!P} = (-3.5 \pm 5.0(stat) \pm 0.2(syst))\times 10^{-2}$, 
showing no sign of CP violation. 
A second CP violating variable is constructed from the positive-negative asymmetry of the triple products
 $C_{\hat T} =\vec{p}_{\mu^\pm} \cdot \vec{p}_p \times \vec{p}_{K^+}$; comparing the positive and negative muon
CP-odd and P-odd observables are obtained. 
The results for the CP-odd observable are $a_{C\!P}^{\widehat{T}-odd} = (1.2 \pm 5.0(stat) \pm 0.7(syst))\times
10^{-2}$, showing again no sign of CP violation, in agreement with SM predictions~\cite{Paracha:2014hca,Alok:2011gv}.

\section{Conclusions}

A very brief account of recent rare decays searches and measurements, 
in the field of \btoqellell transitions was given in this contribution. 
The topic is among the most exiting in current particle physics as small discrepancies 
that point towards possible NP contributions beyond the SM are accumulating.
The results based on the analysis of Run 1 LHC data by the LHCb, 
ATLAS and CMS experiments show tantalising hints of possible deviations
from the SM, especially in the muonic vector coupling. 
These discrepancies will be possibly confirmed or disproved by Run 2 data and through different observables. 
Finally, the LHC experiments will face soon the healthy competition of the Belle II experiment 
in the run for new physics.

\section*{References}

\bibliography{main}

\end{document}